\documentclass[twocolumn,prl,aps]{revtex4}

\usepackage{graphicx}

\begin{document}
\title{Shock waves in ultracold Fermi (Tonks) gases}
\author{Bogdan Damski}
\affiliation{
Instytut Fizyki, Uniwersytet
Jagiello\'nski,
ul. Reymonta 4, PL-30-059 Krak\'ow, Poland}

\begin{abstract}
It is shown that a broad density perturbation in a Fermi (Tonks) cloud 
takes a shock wave form in the course of time
evolution. 
A very accurate analytical description of shock formation is provided.
A simple  experimental setup for the observation of shocks  is discussed.

\end{abstract}

\maketitle
Shock waves show up in various physical systems. For example,
we find them in a  gas bubble driven acoustically,
in  a photonic crystal,
and even in a Bose-Einstein condensate \cite{shw}. In this Letter
we  study properties of 
shock waves in  ultracold Fermi (Tonks) gases: 
the simplest many-body quantum system.
To this aim, we consider dynamics of a Gaussian-like
density perturbation being initially at rest on the fermionic cloud.
We show that such a perturbation 
divides into two pieces that travel
in opposite directions in the course of free
time evolution. The two
perturbations change their shape during propagation, and 
eventually  two shock wave fronts  are formed \cite{landau}.
We provide a detailed theoretical description of their formation 
and further propagation.

Experimental  studies of such phenomena 
are possible due to recent progress in cooling and trapping of fermionic
gases \cite{fer1}. What's more, it turns out that interparticle interactions
between cold trapped  fermions are negligible \cite{fer1}, making the
theoretical description   exactly tractable.

Another exactly solvable many-body system is related to one-dimensional (1D)
hard-core bosonic gas, the so called Tonks gas
\cite{bfmt,olshani2,wrightint}.
As suggested by Olshanii {\it et al.} \cite{olshani2} tightly confined alkali 
atoms are
very promising candidates for experimental realization of the Tonks gas.

There is a close connection between Tonks and 1D Fermi gases due 
to the Fermi-Bose mapping theorem (FBMT) \cite{bfmt,wrightint}.  
This theorem says that if $\Psi(x_1,\cdots,x_N)$ is the ground state 
of the $N$-fermion system described by  (dimensionless) Hamiltonian
\begin{equation}
\label{H1}
\hat{H}= \sum_{i=1}^N\left[-\frac{1}{2} \frac{\partial^2}{\partial x_{i}^{2}} +
V(x_i)\right],
\end{equation}
then the exact many-body ground state  Tonks wave function, in an external
potential the same as in (\ref{H1}), is
$\protect{\prod_{1\le j\le k\le N} {\rm sign}(x_k-x_j) \Psi(x_1,\cdots,x_N)}$
\cite{bfmt,wrightint}. As a result,  the single-particle density
\begin{equation}
\label{ro1}
\varrho_{N}(x)= \int dx_2\cdots dx_N|\Psi(x,x_2,\cdots,x_N)|^2
\end{equation}
is identical for  Fermi and Tonks clouds. Since FBMT can be 
extended to time dependent problems \cite{wrightint}, 
density perturbations on Fermi and Tonks clouds  propagate 
in {\it exactly} the same way.

It was proposed by Kolomeisky and Straley that quantitative  features 
of Tonks (Fermi) density profile can be correctly addressed by the
mean-field approach \cite{ks}, where density
$\varrho_{N}(x)\approx|\Phi(x)|^2$ with $\Phi(x)$ being
the ground state solution of  
\begin{equation}
\label{k1}
-\frac{1}{2}\frac{\partial^2}{\partial x^2}\Phi(x) + V(x)\Phi(x) +
\frac{G}{2} |\Phi(x)|^4\Phi(x)=\mu\Phi(x),
\end{equation}
with $G=\pi^2 N^2$, $N\gg1$, and $|\Phi(x)|^2$ normalized to unity. 
On the basis of (\ref{k1}) we find a distance over which the density 
profile tends to its bulk value $\rho$ when subjected to a localized perturbation.
This characteristic length scale is called the healing length 
$\xi(\rho)$, and equals $1/\sqrt{G}\rho$.
A natural extension of (\ref{k1}) allows for time dependent studies
according to
\begin{equation}
\label{k2}
i\frac{\partial}{\partial t}\Phi(x,t)=\left(
-\frac{1}{2}\frac{\partial^2}{\partial x^2}+ V +
\frac{G}{2} |\Phi|^4\right)\Phi(x,t).
\end{equation}
Therefore, a many-body character of the problem has been overcome
by introducing  nonlinearity into the  theoretical description. 

In this Letter, we investigate formation of shocks and their dynamics
using  both the 
mean-field approach (\ref{k2}) and the exact many-body treatment 
\cite{calculations}.
We show that the mean-field formalism allows for accurate
{\it analytical} determination of all the
quantities concerning shock formation. 
Explicit comparison 
of the mean-field results to the exact calculations allows us to 
specify the range
of applicability of Eq. (\ref{k2}). After 
shock creation the mean-field approach becomes inaccurate, 
and our discussion is based solely on exact findings.

To begin, the dynamics of an initial density 
perturbation in a 1D system will be described. 
This perturbation can
be produced  by an adiabatic focusing of an  appropriately
detuned  laser beam on the particles \cite{pethick}. 
Alternatively, one may cool the cloud 
in the presence of a laser beam \cite{kettfale}.
We assume that our system consists of a zero temperature 
Fermi (Tonks) gas. Atoms are placed 
in a 1D periodic  box having boundaries at $x=\pm l$.

In the Thomas-Fermi regime, when the kinetic energy term in (\ref{k1})
is negligible, 
density profile takes form
\begin{equation}
\label{dens0}
|\Phi(x,0)|^2=\rho_0\sqrt{1+2\alpha e^{-x^2/2\sigma^2}},
\end{equation}
where we assume that the laser produces  potential $u_0\exp(-x^2/2\sigma^2)$,
with $\xi(\rho_0)\ll\sigma\ll l$, and
$\alpha\propto u_0$. We determine $\rho_0$ from the condition
$\int dx |\Phi(x,0)|^2=1$. 
For $\alpha\le0.5$ density profile  can be rewritten as follows
\begin{equation}
\label{dens1}
|\Phi(x,0)|^2\approx \rho_0+ \rho_0(\sqrt{1+2\alpha}-1) e^{-x^2/2\kappa^2},
\end{equation}
where  $\kappa=\sigma\int dx (\sqrt{1+2\alpha e^{-x^2}}-1)/(\sqrt{1+2\alpha}-1)\sqrt{\pi}$ provides a correct normalization of (\ref{dens1}). 

Suppose that the laser beam has been abruptly turned off and the 
perturbation (\ref{dens1}) starts to evolve.
If the system was  described by a single-particle 
Schr\"odinger equation, the perturbation would spread out. 
To find out  what happens in a many-body problem first we consider 
the low amplitude limit.
The ``wave function'' is written as 
$\Phi(x,t)= e^{-i\mu t}[\Phi_0+ \delta\Phi(x,t)]$, where $\Phi_0=1/\sqrt{2l}$
and $\mu=N^2\pi^2/8l^2$. The density profile
reads
\begin{equation}
\label{dens2}
|\Phi(x,t)|^2= \Phi_0^2+
\Phi_0[\delta\Phi(x,t)+\delta\Phi^{\star}(x,t)]+O(|\delta\Phi|^2).
\end{equation}
For $\alpha\rightarrow0$ we find, with the help of (\ref{dens0}) and
(\ref{dens2}), that
\begin{equation}
\label{dens3}
\delta\Phi(x,0)=\alpha\Phi_0(e^{-x^2/2\sigma^2}-\sigma\sqrt{2\pi}/2l)/2+
O(\alpha^2).
\end{equation}
The linearization of (\ref{k2}) leads to the equation 
$$
i\frac{\partial}{\partial t}\delta\Phi= 
-\frac{1}{2} \frac{\partial^2}{\partial x^2}\delta\Phi + G \Phi_0^4(\delta\Phi+\delta\Phi^\star),
$$
which can be solved by means of Bogoliubov substitution:
$\delta\Phi(x,t)= \sum_na_n[{\rm u_n}(x) e^{-i\omega_n t}+ {\rm v}_n^\star(x) e^{i\omega_n t}]$.
Straightforward calculation determines modes ${\rm u}_n$, ${\rm v}_n$
and frequencies $\omega_n$. Coefficients $\{a_n\}$ are found from the initial
condition (\ref{dens3}). Substitution of $\delta\Phi(x,t)$ into (\ref{dens2})
gives
\begin{eqnarray} 
\label{dens4}
|\Phi(x,t)|^2&=&\Phi_0^2+ \alpha\sigma\Phi_0^2\sqrt{2\pi}
\\&&\sum_{n=1}
e^{-k_n^2\sigma^2/2}\cos(k_n x) 
\cos(\omega_n t)/l
+ O(\alpha^2)\nonumber,
\end{eqnarray} 
where $\omega_n=k_n\sqrt{c_0^2+k_n^2/2}$, $c_0=\sqrt{G}\Phi_0^2$, and
$k_n=n\pi/l$. Approximating $\omega_n\approx k_nc_0$ ($\sigma\gg1/c_0$), 
and changing summation into integration we arrive at 
\begin{eqnarray}
\label{dens5}
|\Phi(x,t)|^2 &\approx& \Phi_0^2-\alpha\Phi_0^2\sqrt{2\pi}\sigma/2l
+ \alpha\Phi_0^2[e^{-(x-c_0t)^2/2\sigma^2} 
\nonumber \\&+&  e^{-(x+c_0t)^2/2\sigma^2}]/2
+ O(\alpha^2).
\end{eqnarray}
Thus, we see that  
the initial density profile breaks into two separate pieces
moving with the  sound velocity $c_0$ in the opposite directions.
The dynamics of these perturbations will the subject of subsequent
considerations. 
Finally, notice that the whole discussion applies to systems with $N\gg1$ 
(see (\ref{k1})).

Having in mind the low amplitude result, a hydrodynamical 
approach, that fully accounts for  nonlinear character of the 
problem, will be developed. We write ``wave function'' $\Phi$ as 
$\sqrt{\rho}\exp(i\chi)$ and define the velocity field: $v=\partial_x\chi$.
Eq. (\ref{k2}) in $(\rho,v)$ 
variables for a homogeneous system takes form
\begin{equation}
 \label{hy1}
 \frac{\partial \rho}{\partial t} + \frac{\partial}{\partial x}\left(v\rho\right)
 =0,
\end{equation}
\begin{equation}
\label{h2}
 \frac{\partial v}{\partial t} + \frac{\partial}{\partial x}\left(\frac{1}{2} v^2\right)+\frac{\partial}{\partial x}
 \left(\frac{G}{2}\rho^2-\frac{1}{2}\frac{\partial_x^2 \sqrt{\rho}}{\sqrt{\rho}}\right) = 0,
 \end{equation}
 where (\ref{hy1}) is a continuity equation, while (\ref{h2}) is similar to 
 the Euler equation from classical hydrodynamics. To simplify the problem we
 will neglect in (\ref{h2})  the so called quantum pressure (QP) term 
 $\sim\partial_x^2\sqrt{\rho}/\sqrt{\rho}$ \cite{pethick}. 
 Simple estimation shows that it is of the order 
 of $1/s^2$ with $s$ being the typical spatial scale of variations of
 the density profile. Therefore, it
 is important when $\rho^2G\sim1/s^2$, which means that $s$ is of the order
 of the healing length $\xi$. We assume that initially a perturbation is broad:
 $s\gg\xi$.  
\begin{figure}
\centering
\includegraphics[scale=0.275,clip=true]{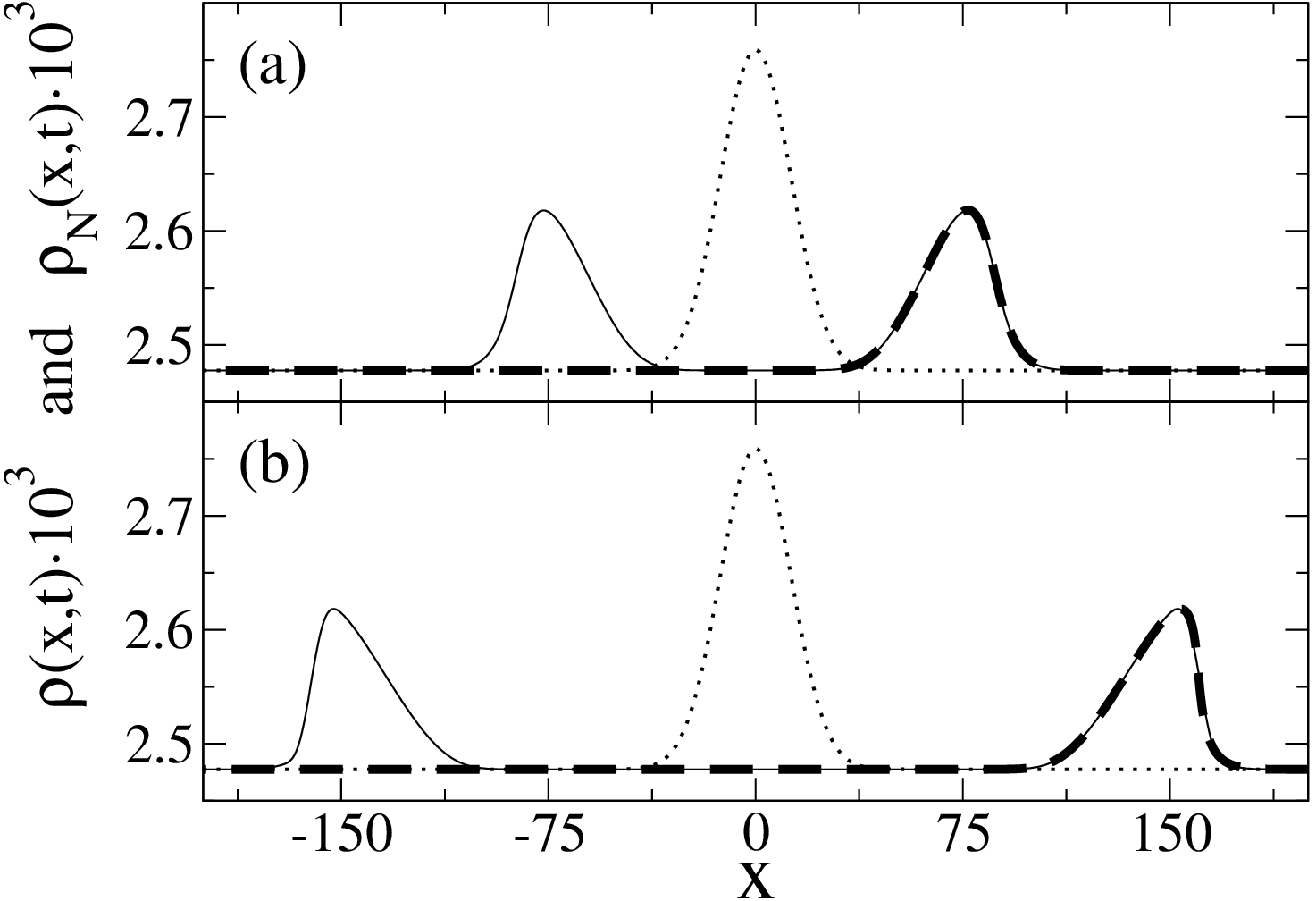}
\caption{
Density of atoms as a function of time. Dotted line 
corresponds to $\rho_{N}(x,0)$.
Solid line is the exact $N$-fermion
solution \cite{calculations}, 
and a dashed line is the analytical approximation (\ref{an1}). 
Plot (a) [(b)] is for $t=$ $22.2$ [$44.4$].
Parameters: $\alpha=0.12$, $u_0\approx-1.16$, $\sigma=12.5$, 
$\kappa\approx12.7$, $N=399$, $l=200$, see \cite{uni} for units.
}
\label{fig1}
\end{figure}

Solution of a similar problem in classical
hydrodynamics is known for years \cite{landau}. Following \cite{landau}
we find exact traveling wave  solutions of 
(\ref{hy1}) and (\ref{h2}) (without the QP term)
$$
\rho=f(x-(a_0\pm2\sqrt{G}\rho)\,t) \ \ \ , \ \ \
v=a_0\pm\sqrt{G}\rho,
$$
where $f$ (an arbitrary function) and $a_0$ (a constant) can be found from
initial conditions. The sign $\pm$ determines a direction
of propagation. This solution is qualitatively the same as the one 
obtained for the  ideal  1D gas \cite{landau}, where the existence of
shocks is  well known. 

We aim at the determination of  dynamics of a single  Gaussian-like
profile moving to the right. Analytical solution of a full problem 
seems to be very difficult  due to the lack of superposition principle. 
On the basis of the low amplitude result (\ref{dens5}),
we assume that initially the right
moving part is a Gaussian of half-amplitude of the initial perturbation 
(\ref{dens1}), and impose a natural condition that $v(x\to\pm l,0)\to0$.
It yields  to
\begin{equation}
\label{an1}
\rho(x,t)=\rho_0+ \rho_0\eta\exp(-[x-\sqrt{G}(-\rho_0+2\rho)\,t]^2/2\kappa^2),
\end{equation}
with $\eta=(\sqrt{1+2\alpha}-1)/2$ (\ref{dens1}). We would like to stress that 
 (\ref{an1}) holds for an arbitrary large $\eta$ since we fully account 
for the nonlinear term. Many interesting results can be obtained from 
(\ref{an1}) despite its implicit form.

First, the  maximum of the impulse has a constant
amplitude equal to $\rho_0(1+\eta)$ and  moves with a  constant velocity ${\cal
V}(\eta)=\sqrt{G}\rho_0(1+2\eta)$. For $\eta\to0$  we get ${\cal
V}_0\approx\sqrt{G}\rho_0=c_0$, which is  the speed of sound propagating 
at the  background density $\rho_0$. In this limit we recover (\ref{dens5}).

Second,  bright perturbations ($\eta>0$) move with higher velocity, 
${\cal V}(\eta)$, than dark structures ($\eta<0$).  
This simple prediction could be directly verified experimentally.
Moreover, it is in a qualitative  agreement with numerical simulations of 
Karpiuk {\it et al.} \cite{karpiuk,karpiuk1}, even though
they  generate excitations in  a
quite different way.

Third, the shape of the perturbation changes in the course of time
evolution -- Fig. \ref{fig1}.
Since the speed of propagation increases with density, the upper part of 
bright impulse moves faster than its tail leading to self-steepening 
in the direction of propagation. Eventually, shock will be formed, i.e.
$\partial \rho(x)/\partial x=-\infty$ in at least one point of the profile.

To determine time and position of shock formation  we solve the  set of equations
\cite{landau}
\begin{equation}
\label{an2}
\frac{\partial x(\rho)}{\partial \rho}= 0 \ \ \ ,  \ \ \
\frac{\partial^2 x(\rho)}{\partial \rho^2}= 0.
\end{equation}
The first one is equivalent to  $|\partial \rho(x)/\partial x| =\infty$, 
and the second one is necessary to assure uniqueness 
of $\rho(x)$. For the density profile (\ref{an1}) one easily finds
$x(\rho)$ and solves (\ref{an2}). Without the loss of generality we assume
from now on
that
$\eta>0$, and find time $t_s$ of shock formation, and density $\rho_s$
at which shock appears 
\begin{equation}
\label{an4}
t_s= \frac{\kappa}{2\sqrt{G}\rho_0\eta e^{-1/2}}\ \ \ , \ \ \
\rho_s=\rho_0+\rho_0\eta e^{-1/2}.
\end{equation}
As a result, an initially broad Gaussian-like density perturbation moving 
in the Fermi (Tonks) cloud (\ref{an1}) 
develops a shock wave front during time evolution.
This is the central result of our paper. The  solution (\ref{dens5})
fails to predict such behavior, since it is derived in the zero amplitude 
limit ($\eta\to0$) where $t_s\to\infty$.   

Interestingly, the time of shock formation can be correctly
estimated without solving  (\ref{an2}). 
Indeed, the half-width of the impulse ($\approx 2\kappa$)
is a difference in a distance traveled by 
the  upper and lower parts of the impulse 
until  shock appears: $2\kappa\approx({\cal V}(\eta)-{\cal V}_0)t_s$. It 
gives $t_s\sim\kappa/\sqrt{G}\eta\rho_0$. 
Estimating the position of a shock wave front by position of impulse's
maximum we get a very simple expression: $x_s= \kappa e^{1/2}(1+1/2\eta)$.

\begin{figure}
\centering
\includegraphics[width=8.6cm,clip=true]{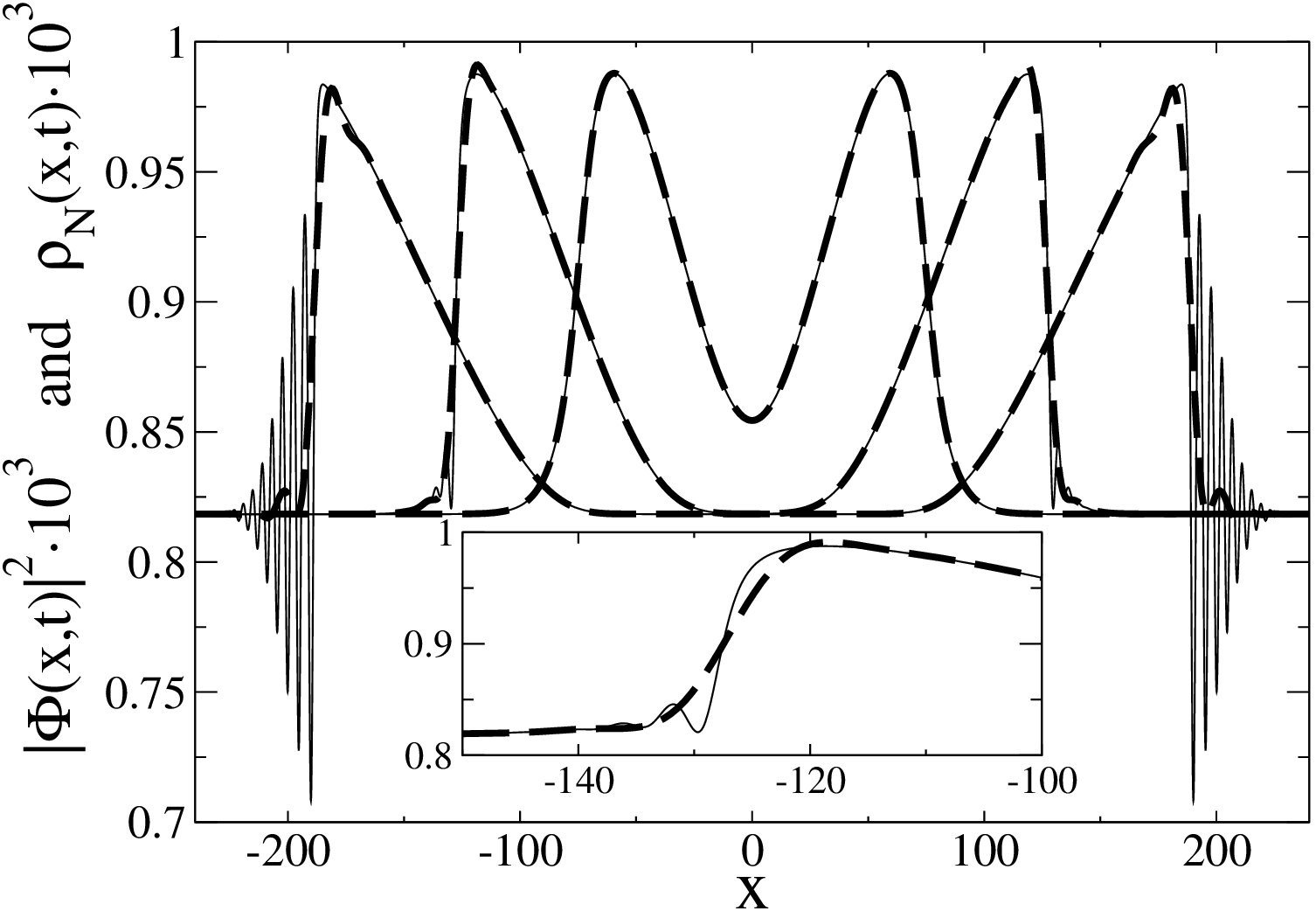}
\caption{
Density of atoms as a function of time. Subsequent profiles correspond to
$t=41$, $82$, $131.2$.
Dashed line is an exact $N$-body calculation \cite{calculations}, 
while  solid line
presents numerical solution of mean-field equation (\ref{k2}).
Inset shows details of a density profile at $t=82$. Shock's characteristics:
$t_s=81.6$ and $x_s=118.4$. Parameters:
$\alpha=0.5$, $u_0\approx-0.53$, $\sigma=20$, $l=600$, $N=399$, 
$\xi(0.8\cdot10^{-3})\approx1$, see \cite{uni} for units.
}
\label{fig2}
\end{figure}

Comparison of the analytical solution (\ref{an1}) to the $N$-body
one \cite{calculations} for a moderately large initial perturbation ($x_s>l$)
shows excellent agreement--see
Fig. \ref{fig1}. As $x_s$ becomes smaller than half-box-size ($=l$)
shock waves show up in our simulations.  
We have increased the amplitude of the
initial perturbation and compared a full mean-field calculation  
with the exact fermionic one \cite{calculations}.
As shown in Fig. \ref{fig2}, until the expected moment of shock wave 
creation agreement is very good. The analytical solution (\ref{an1}),
not presented in Fig. \ref{fig2}, fits nicely to exact calculation as long as
$t<t_s$. Then, the mean field approach
fails to reproduce exact calculations at the {\it front} edge, and non-physical atom 
oscillations appear. 
We found that the QP term, playing a role when a spatial scale of 
density variations ($s$) becomes comparable to the healing length $\xi$,
 is responsible for their appearance. 
As $t$ tends to $t_s$ we find  $s\to\xi$ (Fig. \ref{fig2}).
Therefore, we find that  the dynamical mean-field approach (\ref{k2})
works nicely as long as $s\gg\xi$ and fails when $s\sim\xi$. In this way
we bring  a quantitative criteria into discussion \cite{dyskusja,wrightint}
concerning applicability of the Kolomeisky and Straley equation (\ref{k2})
to Tonks (Fermi) systems.

Although, the derivative of density does not exactly
go to infinity, the density profile becomes very steep at the front edge --
Fig. \ref{fig2}.
To look more quantitatively at  shocks' dynamics 
we define the symmetry coefficient $\Lambda$ as 
$\int_{x_{m}}^ldx[\rho_N(x)-\rho_N(l)]/\int_0^{x_{m}}dx[\rho_N(x)-\rho_N(l)]$, i.e.
as a ratio between the number of atoms in the front and rear impulse's parts
with respect to the impulse maximum being at $x_m>0$.
$\Lambda$ is close to zero for
a well developed shock wave profile, and equals
to one for symmetric perturbations.

\begin{figure}
\centering
\includegraphics[scale=0.275,clip=true]{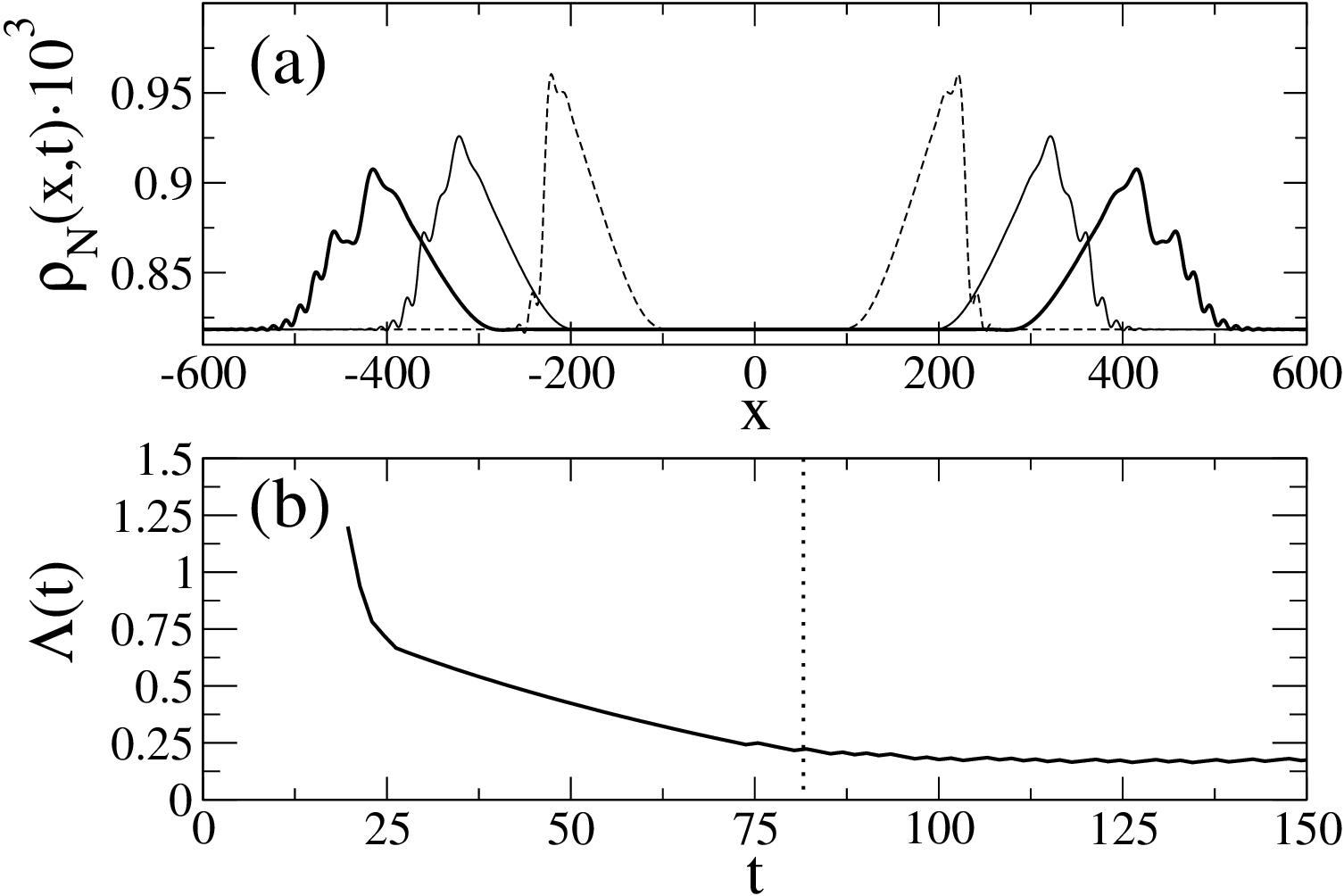}
\caption{
Plot (a): further time evolution of the density profile from Fig. \ref{fig2}.
Subsequent profiles correspond to $t=164$, $262.4$, $344.4$.
Plot (b): symmetry coefficient $\Lambda(t)$. Dotted line
shows  expected moment of the shock formation. Both plots present exact 
$N$-body results \cite{calculations}. Parameters:  as in  Fig. \ref{fig2},
see \cite{uni} for units.
}
\label{fig3}
\end{figure}

In the first stage of evolution, the profile undergoes self-steepening 
dynamics so that $\Lambda$ decreases reaching the lowest value approximately
at 
$t=t_s=81.6$ -- Fig. \ref{fig2}, \ref{fig3}b. 
Then $\Lambda(t)$ decreases a little and  becomes flat --
impulse propagates roughly without change of shape 
until 
$t\approx150$ (Fig. \ref{fig2}, \ref{fig3}b).
In the course of further evolution a shock wave front
gradually disappears
and the density profile becomes more and more symmetric ($\Lambda\to1$) 
-- Fig. \ref{fig3}a. Therefore, 
we conclude that there are three stages of  evolution: the shock wave
formation, propagation of a shock-like impulse, and the impulse 
{\it explosion}
leading to broadening of the  density profile.

\begin{figure}
\centering
\includegraphics[scale=0.275,clip=true]{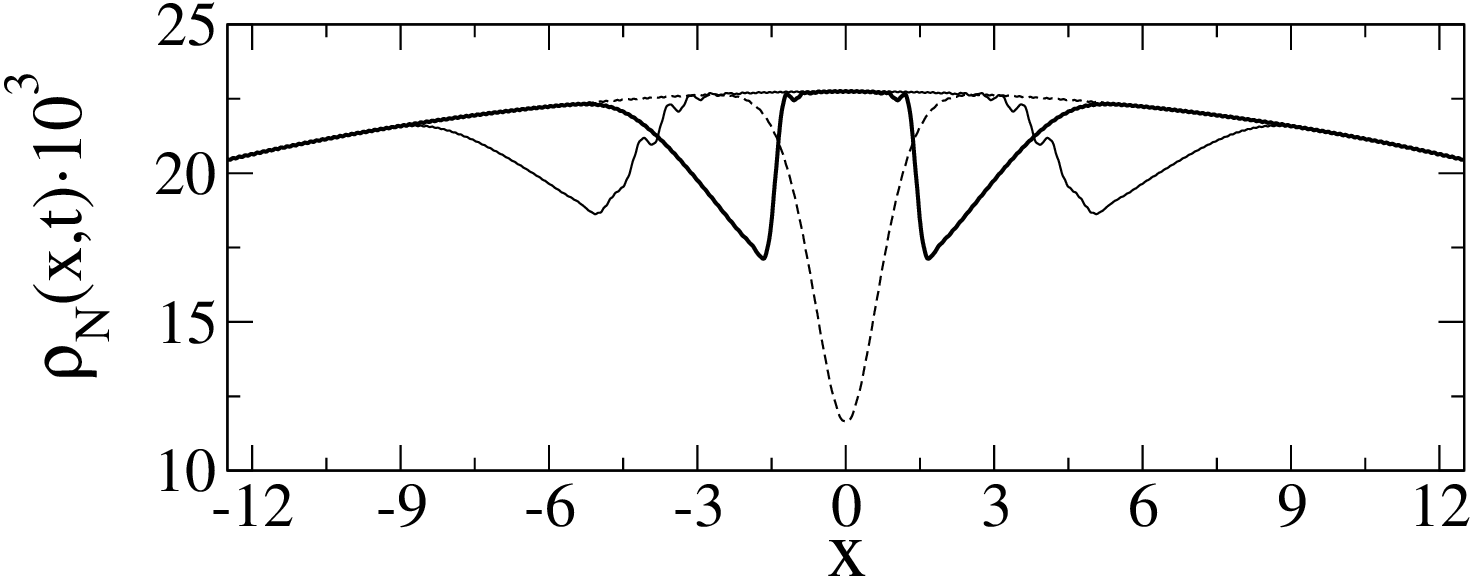}
\caption{The exact many-body time evolution of the density profile of $N=399$
fermions in a harmonic trap \cite{calculations}. Subsequent profiles correspond to
 $t=0$, $4\pi/120$, $9\pi/120$ ($2\pi$ is a trap period) \cite{units}. 
Notice that
for dark perturbations  back edge self-steepens instead of the front one.
Parameters: $\sigma=0.75$, $u_0=300$. 
}
\label{fig4}
\end{figure}

Finally, let us comment on the dynamics of density perturbations
 in a harmonically trapped case:  $V(x_i)=x_i^2/2$ (\ref{H1}).
The division of an initial density perturbation into two similar pieces
happens  independently of the location of the laser beam focus.
Just after the laser turn off the two perturbations travel in opposite directions
to the edge of the cloud and come back to join after half-trap-period. At this
moment the density profile is a mirror image of the initial shape. Then
another splitting takes place and the dynamics repeats itself. 
As in the homogeneous case, perturbations undergo self-steepening dynamics leading
to formation of a shock-like front, which then blows up -- Fig. \ref{fig4}.

Quasi-one-dimensional atomic traps form  very
promising systems for observation of shock waves. 
Indeed, their reduced dimensionality 
avoids dispersion of the impulse amplitude due to  
multidimensional propagation.
The periodicity of shocks  formation in a harmonic trap should
allow for their multiple {\it in situ} 
observations \cite{kettfale}
in a single run of an experiment. 

It is worthwhile to point out that the described  splitting of the initial 
perturbation has been already experimentally observed 
in a Bose-Einstein condensate (BEC) \cite{kettfale}. An extension 
of our calculations to Bose gases provides a theoretical
description of a shock formation in a BEC. 
These results are  presented in \cite{bodzio}.

To summarize, we have studied in detail dynamics of density 
wave packets in the Fermi (Tonks) cloud. 
We have derived a simple analytical expression for 
the speed of propagation of Gaussian-like 
density wave packets having {\it arbitrary} amplitude. 
This result can be crucial for determination of the sound
velocity from experimental data. 
We have also described analytically the 
process of a shock wave formation, and proposed a simple 
experimental setup for its observation.
This prediction can be {\it directly} verified in a single run of
experiment. 
Finally, we have found that 
the ``quantum'' shock wave front blows up at some point instead of becoming
more and more steep in the course of free time evolution.  
The presence of shock waves in  three-dimensional configurations 
can be an interesting subject for future studies.
Indeed, as pointed out in \cite{karpiuk1}, the propagation of density
perturbations in a three-dimensional Fermi gas can differ significantly  
from a quasi-one-dimensional case. The hydrodynamical approach proposed in 
\cite{karpiuk1} supported by exact many-body calculations 
can be used for such investigations.

I would like to acknowledge 
discussions with M. Brewczyk, Z. P. Karkuszewski, K. Sacha, and J. Zakrzewski. 
Work supported by KBN project 2 P03B 124 22.


\begin{thebibliography}{99}
\bibitem{calculations} We find the 
$N$ lowest eigenstates of the  Hamiltonian 
$-\frac{1}{2}\frac{\partial^2}{\partial x^2} + V(x)+ u_0\exp(-x^2/2\sigma^2)$
[$V(x)=$ $0$ or $x^2/2$]. After laser turn off ($u_0=0$) they
 undergo an ordinary single-particle time evolution. The many-body Fermi
wave function  is a Slater 
determinant built out of them.
\bibitem{uni} 
Since we consider  here a homogeneous system, we 
take arbitrary length scale $l_0$ for a unit of length.
For a unit of time we choose $ml_0^2/\hbar$,
where $m$ is the atomic mass.
\bibitem{units} 
We use here the harmonic oscillator units where a unit of length equals
$\sqrt{\hbar/m\omega}$ and a unit of time is $1/\omega$. Naturally,
$\omega$ is a trap frequency and $m$ is an atomic mass. For example,
taking  a realistic case of $\omega=2\pi\cdot10$Hz,
density profiles in  Fig. \ref{fig4}
correspond to $t=0$, $1.7$ms, $3.8$ms.
\end{thebibliography}
\end{document}